\documentclass[conference]{IEEEtran}
\IEEEoverridecommandlockouts
\usepackage{booktabs} 
\usepackage{stfloats}
\usepackage{cite}
\usepackage{amsmath,amssymb,amsfonts}
\usepackage{hyperref}
\usepackage{algorithmic}
\usepackage{graphicx}
\usepackage{textcomp}
\def\BibTeX{{\rm B\kern-.05em{\sc i\kern-.025em b}\kern-.08em
    T\kern-.1667em\lower.7ex\hbox{E}\kern-.125emX}}
\begin{document}

\title{VARTS: A Tool for the Visualization and Analysis of Representative Time Series Data\\
}

\author{\IEEEauthorblockN{Duosi Jin}
\IEEEauthorblockA{\textit{College of Computer Science}\\
\textit{ and Technology} \\
\textit{Nanjing University of Aeronautics}\\
\textit{  and Astronautics}\\
Nanjing, China \\
jinduosi@nuaa.edu.cn}
\and
\IEEEauthorblockN{Jianqiu Xu\thanks{\textsuperscript{*} Corresponding author. The authors acknowledge the support of the National Science Foundation under grants Nos.(U23A20296, 62472217).}\textsuperscript{*}}
\IEEEauthorblockA{\textit{College of Computer Science}\\
\textit{  and Technology} \\
\textit{Nanjing University of Aeronautics}\\
\textit{  and Astronautics}\\
Nanjing, China \\
jianqiu@nuaa.edu.cn}
\and
\IEEEauthorblockN{Guidong Zhang}
\IEEEauthorblockA{\textit{School of Automation} \\
\textit{Guangdong University of Technology}\\
Guangdong, China \\
guidong.zhang@gdut.edu.cn}
}

\maketitle

\begin{abstract}
Large-scale time series visualization often suffers from excessive visual clutter and redundant patterns, making it difficult for users to understand the main temporal trends. To address this challenge, we present VARTS, an interactive visual analytics tool for representative time series selection and visualization. Building upon our previous work M4-Greedy, VARTS integrates M4-based sampling, DTW-based similarity computation, and greedy selection into a unified workflow for the identification and visualization of representative series. The tool provides a responsive graphical interface that allows users to import time series datasets, perform representative selection, and visualize both raw and reduced data through multiple coordinated views. By reducing redundancy while preserving essential data patterns, VARTS effectively enhances visual clarity and interpretability for large-scale time series analysis. The demo video is available at https://youtu.be/mS9f12Rf0jo.
\end{abstract}


\section{Introduction} 

The rapid development of sensing, monitoring, and data acquisition technologies has led to massive collections of time series data across domains such as aviation, manufacturing, and the Internet of Things. 
Visual analytics plays a crucial role in enabling users to explore temporal patterns, detect anomalies, and derive insights from such data. 
However, when visualizing large-scale time series datasets, users often encounter two major challenges: 
(i) the overwhelming number of points within each series causes severe visual clutter and performance degradation, and 
(ii) displaying numerous correlated series simultaneously tends to obscure the underlying temporal trends.
\begin{figure}[htbp]
    \centering
    \centerline{\includegraphics{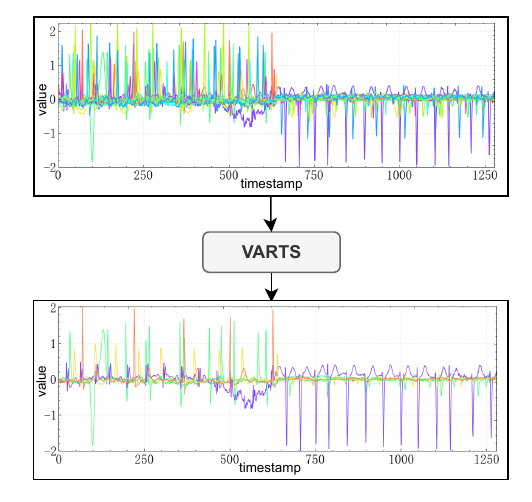}}
    \caption{Comparison of raw and representative time series visualization. 
    VARTS automatically selects and visualizes a concise subset that preserves the main patterns while reducing visual clutter.}
    \label{fig-intro}
    \vspace{-1em}
\end{figure}

To address these challenges, our previous works have explored both visualization frameworks and data reduction techniques. 
In our earlier system, \textit{FlyTsViz}~\cite{Jin2025FlyTsViz}, we developed a flight time series visualization platform based on data reduction methods, 
which enabled interactive exploration of large-scale flight datasets. 
Building upon this foundation, we later proposed the \textit{M4-Greedy} algorithm~\cite{Jin2025M4Greedy}, a visualization-oriented representative selection approach that jointly optimizes diversity and coverage among selected time series. 
By combining M4-based temporal sampling with DTW-based greedy selection, 
the algorithm efficiently identifies representative time series subsets that preserve global shape characteristics while reducing data volume by up to an order of magnitude. 
Although the algorithm demonstrates strong quantitative performance, 
users still lack an intuitive tool to apply, observe, and interactively explore its selection process in real-world analytical scenarios.

In this paper, we present \textbf{VARTS} (short for \textit{Visualization and Analysis of Representative Time Series Data}), an interactive visual analytics tool that integrates the M4-Greedy representative selection algorithm into an enhanced visualization platform for large-scale time series data. 
The tool is implemented in Qt, providing a responsive graphical interface that allows users to upload datasets, perform representative selection, and visualize both raw and reduced data through multiple coordinated views. VARTS bridges algorithmic research and practical application by transforming a theoretical selection framework into an interactive, extensible, and domain-oriented visual analytics tool.

The tool is demonstrated on real-world flight time series datasets, 
where users can interactively adjust parameters such as the number of representatives ($k$) and the diversity-coverage trade-off ($\alpha$) 
to explore data behavior at different levels of abstraction. 
VARTS effectively reduces visual clutter while preserving key temporal patterns, 
facilitating faster and clearer understanding of large-scale time series data.
As illustrated in Fig.~\ref{fig-intro}, VARTS transforms dense, cluttered raw time series data into a concise and interpretable visualization by automatically selecting representative sequences. 
The upper panel of the figure shows multiple overlapping time series, which are difficult to interpret due to visual congestion. 
After applying the M4-Greedy-based representative selection, the lower panel displays only the most informative series, 
preserving the essential temporal patterns while substantially improving clarity. 
This reduction not only enhances readability but also maintains the overall structural fidelity of the dataset.

The main contributions of this demo are summarized as follows:
\begin{enumerate}
    \item We present VARTS, a tool for the visualization and analysis of representative time series data.
    \item The tool implements the M4-Greedy algorithm to automatically identify and visualize representative subsets, reducing visual clutter and enhancing interpretability.
    \item The demo illustrates how users can interactively explore flight time series data through parameter control and coordinated multi-view visualization.
\end{enumerate}

\section{Related Work} 

\textbf{Time Series Visualization.}
Traditional visualization tools such as Tableau, Power BI, and ECharts provide interactive charting interfaces but are limited in handling large-scale time series data. 
Recent research (e.g., OM3~\cite{ref_OM3}, and IoTDB-based visualization~\cite{ref_M4_LSM}) 
focus on improving scalability and interactivity through hierarchical aggregation or multi-scale representations. 
However, these methods are primarily designed as algorithmic frameworks rather than domain-oriented visual analytics platforms.

\textbf{Representative Selection and Sampling.}
Representative selection has been widely studied to summarize large collections of time series data. 
Early works relied on clustering-based approaches~\cite{ref_K_Shape2016}, 
while more recent studies consider diversity and coverage jointly~\cite{ref_Pan2005,ref_TowardsEfficientSelection,ref_RepresentativeTimeSeries}. 
These methods inspired the M4-Greedy algorithm~\cite{Jin2025M4Greedy}, 
which jointly optimizes diversity and coverage under the DTW (Dynamic Time Warping) metric.
Similar greedy selection strategies have also been adopted in related fields such as spatial crowdsourcing~\cite{ref_Tong2016Online}.

\textbf{M4-Greedy Algorithm and Tool Integration.}
Our previous work, FlyTsViz~\cite{Jin2025FlyTsViz}, presented a time series visualization tool focusing on reduction and visual analytics, 
while M4-Greedy~\cite{Jin2025M4Greedy} introduced an algorithmic framework for representative selection. 
VARTS integrates these two lines of work by embedding the M4-Greedy algorithm into a practical visualization environment, 
bridging computational optimization and interactive exploration.

\section{The Framework}

In addition to the representative selection module, VARTS also supports general visualization and anomaly visualization. 
However, this paper primarily focuses on the representative selection mechanism and its integration with the visualization workflow.

\begin{figure}[htbp]
\vspace{-1.5em}
\centerline{\includegraphics{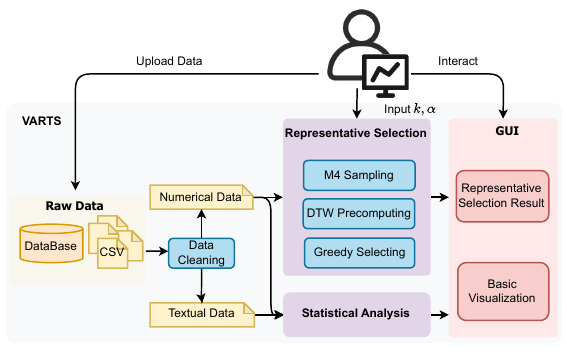}}
\caption{The framework of VARTS.}
\label{fig-framework}
\vspace{-1em}
\end{figure}

The overall framework of VARTS is illustrated in Fig.~\ref{fig-framework}. 
It integrates data preprocessing, representative selection, and interactive visualization within a unified workflow. 
Specifically, the framework consists of three major components: 
\textbf{(i) Data Preprocessing}, 
\textbf{(ii) Representative Selection}, 
and \textbf{(iii) Visualization and Interaction}. 
The following subsections describe each component in detail.
\subsection{Data Preprocessing}

VARTS accepts multiple data formats, including databases, CSV files, and text-based records, which contain both numerical and textual attributes.
After data upload, the tool performs cleaning, normalization, and type classification to separate numerical and categorical variables. 
In addition, missing values and inconsistent timestamps are automatically handled to maintain data integrity and temporal consistency.

\subsection{Representative Selection}

The representative selection part is the core module of VARTS, responsible for reducing data redundancy and preserving the overall temporal patterns of large-scale time series datasets. 
This part integrates three algorithmic components, namely \textit{M4 Sampling}, \textit{DTW-based Similarity Computation}, and \textit{Greedy Representative Selection}, to automatically identify a compact subset of time series that best capture the global characteristics of the dataset.

\subsubsection{M4 Sampling}
To efficiently describe the shape of each time series, VARTS employs the M4 sampling method~\cite{ref_M4}, which extracts four key points from each segment: 
the point with the minimum timestamp, the maximum value, the minimum value, and the point with the maximum timestamp. 
These four points (M4) effectively retain both the amplitude and temporal shape of local subsequences, 
forming a compact yet informative representation. 
This step greatly reduces the computational burden of subsequent similarity computation while maintaining the overall structure of the data.

\subsubsection{DTW-based Similarity Computation}
Given the sampled representations, the tool computes pairwise similarities between time series using the DTW distance.
Unlike Euclidean distance, DTW allows flexible time stretching and compression, 
making it more robust for time series data where phase shifts or local delays may occur due to variations in sensor sampling. 
To improve efficiency, a precomputation and caching mechanism is applied to store intermediate DTW results, 
significantly accelerating repeated similarity queries during the selection process.

\subsubsection{Greedy Representative Selection}

Based on the DTW distance matrix, VARTS applies a greedy strategy to iteratively select representative time series. 
Let $\mathcal{T} = \{T_1, T_2, \dots, T_n\}$ denote the set of $n$ time series, 
and let $D(T_i, T_j)$ represent the DTW distance between two series $T_i$ and $T_j$. 
The goal is to select a subset $\mathcal{R} \subset \mathcal{T}$ of size $k$ that maximizes the representativeness of the dataset by jointly optimizing \textit{diversity} and \textit{coverage}, following the M4-Greedy formulation~\cite{Jin2025M4Greedy}.

\textbf{Objective Function.}
The optimization objective is defined as:
\begin{equation}
\max_{\mathcal{R} \subset \mathcal{T},\, |\mathcal{R}| = k}
    \Big[ \alpha \cdot \text{Div}(\mathcal{R}) - (1-\alpha) \cdot \text{Cov}(\mathcal{R}, \mathcal{T}) \Big],
\label{eq:objective_m4}
\end{equation}
where $\alpha$ is non-negative coefficients balancing the two objectives.

The \textit{diversity} term encourages the selected representatives to be well-separated:
\begin{equation}
\text{Div}(\mathcal{R}) = \min_{T_i, T_j \in \mathcal{R},\, T_i \neq T_j} D(T_i, T_j),
\label{eq:div_m4}
\end{equation}
which maximizes the smallest pairwise DTW distance among representatives.
This max-min formulation prevents redundant selections and ensures that all representatives remain distinct in temporal patterns.

The \textit{coverage} term measures how well the representatives approximate the original dataset:
\begin{equation}
\text{Cov}(\mathcal{R}, \mathcal{T}) = \frac{1}{n} \sum_{T_i \in \mathcal{T}} \min_{T_j \in \mathcal{R}} D(T_i, T_j),
\label{eq:cov_m4}
\end{equation}
where a smaller value indicates that most series are close to at least one representative under the DTW metric.

\textbf{Greedy Optimization.}
Starting from an empty representative set $\mathcal{R} = \varnothing$, 
the algorithm iteratively selects the time series $T^\ast$ that yields the greatest incremental improvement in the composite objective defined in Eq.~\ref{eq:objective_m4}:
\begin{equation}
T^\ast = \arg\max_{T \in \mathcal{T} \setminus \mathcal{R}}
    \Big[ \alpha \cdot \Delta \text{Div}(\mathcal{R}, T) - (1-\alpha) \cdot \Delta \text{Cov}(\mathcal{R}, T) \Big],
\label{eq:greedy_m4}
\end{equation}
where $\Delta \text{Div}(\mathcal{R}, T)$ and $\Delta \text{Cov}(\mathcal{R}, T)$ denote the incremental changes in diversity and coverage when adding candidate $T$ into the current representative set $\mathcal{R}$. 
This greedy process continues until $|\mathcal{R}| = k$, producing a representative subset that maintains high intra-set diversity and low global coverage, 
thereby preserving the essential temporal structure of the dataset while minimizing redundancy in visualization.


\subsection{Visualization and Interaction}
The visualization and interaction component enables users to explore both raw and representative time series data through a graphical interface built on the Qt framework. 
It allows users to visually compare results and assess the outcomes of the representative selection process.

VARTS supports multiple visualization modes for flight time series data, including \textit{line charts}, \textit{box plots}, and \textit{trajectory maps}. 
These coordinated views allow users to examine temporal dynamics, statistical distributions, and spatial trajectories of flight parameters, offering both global and local perspectives on flight behavior. 

After representative selection, the tool displays only the selected series in the visualization views. Users can adjust parameters such as the number of representatives ($k$) and the diversity-coverage trade-off ($\alpha$), with the visualizations updated accordingly. 
This integration between the frontend interface and the backend computation supports interactive exploration of time series data.




In summary, the VARTS framework seamlessly combines data preprocessing, representative selection, and visual interaction, forming a complete workflow for intelligent and scalable time series visual analytics.
\begin{figure*}[htbp]
\centerline{\includegraphics{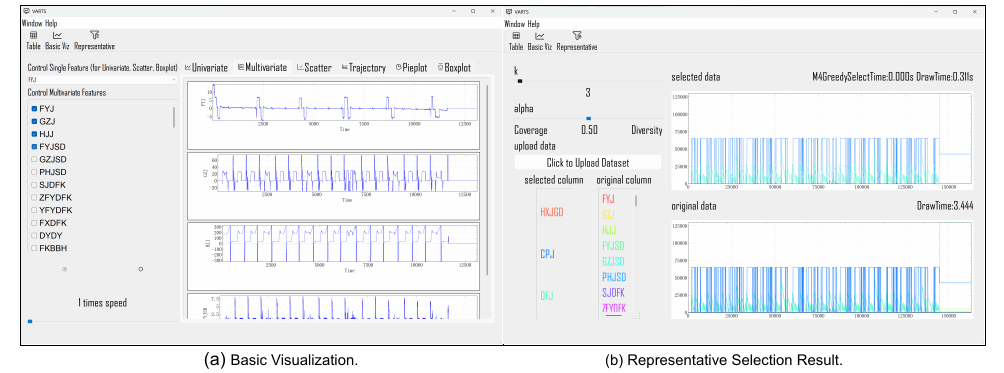}}
\caption{Screenshots of VARTS.}
\label{fig-screenshots}
\vspace{-1em}
\end{figure*}

\section{Demonstration}

Building upon the tool framework described in the previous section, this demonstration presents the interactive workflow of VARTS, which integrates multi-view visualization and representative selection for large-scale time series analysis. Fig.~\ref{fig-screenshots} shows screenshots of VARTS. After uploading a dataset, users can explore the trend of individual time series dimensions and their interrelations in Fig.~\ref{fig-screenshots}(a). In Fig.~\ref{fig-screenshots}(b), an overview of the entire dataset is provided, enabling users to identify which dimensions significantly influence the overall data distribution. For dimensions of interest, users can return to Fig.~\ref{fig-screenshots}(a) to select and closely examine them.
\vspace{-1em}
\begin{table}[htbp]
\caption{Flight Time Series Datasets Used in VARTS}
\vspace{-1em}
\begin{center}
\resizebox{\columnwidth}{!}{
\begin{tabular}{lccc}
\toprule
\textbf{Dataset} & \textbf{\#Tuples$^{\mathrm{a}}$} & \textbf{\#Variables} & \textbf{Size} \\
\midrule
FlightData-2021 & 78181 & 103 & 64 MB \\
FlightData-2023 & 110115 & 86 & 56 MB \\
FlightData-2024 & 545850 & 92 & 298 MB \\
\bottomrule
\multicolumn{4}{l}{$^{\mathrm{a}}$Each tuple corresponds to one time step in the flight time series dataset.} \\
\end{tabular}}
\label{tab-datasets}
\end{center}
\vspace{-1.5em}
\end{table}

\subsection{Data Import and Preprocessing}

Users can load time series data from either databases or local CSV files. 
VARTS automatically parses, normalizes, and aligns the data to ensure temporal consistency. 
Table~\ref{tab-datasets} summarizes the datasets used in the demonstration, which vary in the number of tuples, variables, and overall size, illustrating the scalability of the tool in handling diverse flight data sources. Additionally, the representative selection function can be used independently, provided that the dataset contains multivariate numerical data.

\subsection{Visualization and Representative Selection}

After data import, users can explore the raw time series through coordinated visualizations including \textit{line charts}, \textit{box plots}, and \textit{trajectory maps}, as shown in Fig.~\ref{fig-screenshots}(a). 
These views provide complementary perspectives on temporal variation, statistical distribution, and spatial behavior of flight parameters. 
The representative selection module applies the M4-Greedy algorithm to identify a concise subset of representative series. 
An initial visualization is generated automatically after data loading, where both the original and representative time series are displayed side by side in Fig.~\ref{fig-screenshots}(b). 
This comparative view highlights how the selected subset preserves the dominant temporal patterns of the dataset while substantially reducing visual clutter and redundancy.

\subsection{Interactive Exploration}

VARTS supports real-time interaction for parameter tuning and result comparison. 
Users can adjust the number of representatives ($k$) and the diversity-coverage trade-off ($\alpha$) via sliders located in Fig.~\ref{fig-screenshots}(b): when $\alpha \!\to\! 0$, selection favors coverage, while $\alpha \!\to\! 1$ emphasizes diversity. 
After the DTW similarity matrix is computed once during the initial selection, subsequent adjustments of $k$ or $\alpha$ only update the visualization based on precomputed results. 
Displaying the representative series takes only about 10\% of the time required to render all original time series, as the number of representative dimensions is typically below 10 while the original dataset contains on the order of 100 variables, ensuring smooth and responsive interaction.

This demonstration highlights how VARTS bridges algorithmic analysis and human insight, providing an effective and extensible platform for representative time series visualization in aviation analytics.



\section{Conclusion}

This demo presents VARTS, an interactive visual analytics tool for large-scale time series data. The tool implements the M4-Greedy representative selection algorithm, integrating M4 sampling, DTW-based similarity computation, and greedy selection into a practical visualization workflow. Demonstrated on real flight time series datasets, VARTS enables users to explore and compare raw and representative series, adjust selection parameters, and analyze patterns with reduced visual clutter.


\begin{thebibliography}{00}

\bibitem{Jin2025FlyTsViz}
D. Jin, J. Xu, and X. Zhao, ``FlyTsViz: A Flight Time Series Data Visual Analytics System Based on Reduction Method,'' 
in \textit{Proc. 8th Int. Conf. on Data Storage and Data Engineering (DSDE)}, 2025, pp. 30--36.

\bibitem{Jin2025M4Greedy}
D. Jin, S. Xu, M. Ju, Z. Zhang, and F. Lin, ``M4-Greedy: Visualization-Oriented
Representative Selection of Time Series Data,'' 
in \textit{The 1st Int. APWeb-WAIM Workshop on Spatial-rich Big Data Management and Applications (SRBDMA2025)}, in press.

\bibitem{ref_OM3}
Y. Wang \textit{et al.}, ``OM3: An ordered multi-level min-max representation for interactive progressive visualization of time series,'' 
\emph{Proc. ACM Manag. Data}, vol. 1, no. 2, pp. 145:1--145:24, 2023.

\bibitem{ref_M4_LSM}
L. Rui, X. Huang, S. Song, Y. Kang, C. Wang, and J. Wang, 
``Time Series Representation for Visualization in Apache IoTDB,'' 
\emph{Proc. ACM Manag. Data}, vol. 2, no. 1, pp. 35:1--35:26, 2024.

\bibitem{ref_K_Shape2016}
J. Paparrizos and L. Gravano, ``k-Shape: Efficient and Accurate Clustering of Time Series,'' in \textit{Proc. ACM SIGMOD Int. Conf. Manage. Data}, 2015, pp. 1855--1870.

\bibitem{ref_Pan2005}
F. Pan, W. Wang, A. K. H. Tung, and J. Yang, ``Finding representative set from massive data,'' in \textit{Proc. 5th IEEE Int. Conf. on Data Mining (ICDM)}, sep. 2005. pp. 8--15.

\bibitem{ref_TowardsEfficientSelection}
C. Yang, L. Chen, H. Wang, and S. Shang, ``Towards efficient selection of activity trajectories based on diversity and coverage,'' 
in \textit{Proc. AAAI Conf. on Artificial Intelligence(AAAI)}, 2021, vol. 35, no. 1, pp. 689--696.


\bibitem{ref_RepresentativeTimeSeries}
G. Lee, S. Huang, Z. Bao, and Y. Zhao, ``Representative Time Series Discovery for Data Exploration,'' 
\emph{Proc. VLDB Endowment}, vol. 18, no. 3, pp. 915--928, 2024.

\bibitem{ref_Tong2016Online}
Y. Tong, J. She, B. Ding, L. Wang, and L. Chen, 
``Online mobile Micro-Task Allocation in spatial crowdsourcing,'' 
in \textit{Proc. 32nd IEEE Int. Conf. on Data Engineering (ICDE)}, 2016, pp. 49--60.

\bibitem{ref_M4} 
U. Jugel, Z. Jerzak, G. Hackenbroich, and V. Markl, ``M4: A visualization-oriented time series data aggregation,'' 
\emph{Proc. VLDB Endowment}, vol. 7, no. 10, pp. 797--808, 2014.

\end{thebibliography}
\end{document}